\documentclass{mn2e}
 
\newif\ifAMStwofonts

\input epsf    

\title[Potential--density pairs for spherical galaxies...]{Potential--density
pairs for spherical galaxies and bulges: the influence of scalar fields.}

\author[M. A. Rodr\'\i guez--Meza and Jorge L.
Cervantes-Cota]{M. ~A.~ Rodr\'\i guez--Meza $^{1,2}$
and Jorge ~L. ~Cervantes--Cota$^{1,3}$ \\
$^1$Depto.  de F\'{\i}sica, Instituto Nacional de Investigaciones Nucleares
P.O. Box 18-1027, M\'exico D.F. 11801, M\'exico \\
$^2$Instituto de F\'{\i}sica,  Universidad Aut\'onoma de Puebla
P.O. Box J-48, Puebla 72570, M\'exico \\
$^3$Astrophysikalisches Institut Postdam, An der Sternwarte 16, 
D-14482 Potsdam, Germany}

\begin{document}
\maketitle

\begin{abstract}
A family of potential--density pairs has been found for spherical
halos and bulges of galaxies in the Newtonian limit of scalar--tensor
theories of gravity.   The scalar field is described by a Klein--Gordon
equation with a source that is coupled to the standard Poisson equation of
Newtonian gravity.  The net gravitational force is given by two
contributions: the standard Newtonian potential plus a term stemming from
massive scalar fields.  General solutions have been found for spherical
systems.  In particular, we compute potential--density pairs of 
spherical galactic systems, and some other astrophysical quantities that 
are relevant to generating initial conditions for spherical galaxy simulations.
\end{abstract}

\begin{keywords}
stellar dynamics -- elementary particles -- Galaxy: kinematics and dynamics --
Galaxy: halo-- Galaxy: bulge --  Galaxy: structure
\end{keywords}



\section{Introduction}
In recent years there has been important progress in
understanding the dynamics that led to the formation
of galaxies. Two and three dimensional $N$-body simulations
of galaxies and protogalaxies have been computed using
millions of particles, giving a more realistic view of how galaxies,
quasars, and black holes could have formed \cite{Ba98,BaHe92}.

The composition of Universe at the time galaxies formed could 
theoretically be very varied, including visible and dark baryonic matter,
non--baryonic dark matter, neutrinos, and many cosmological relics
stemming from symmetry breaking processes predicted by high energy
physics \cite{KoTu90}.   All these particles and fields, if present, should
have played a role in  structure formation.  Accordingly,  recent 
independent observational data;  CMBR at various angular
scales \cite{deB00,Be03},  type Ia supernovae \cite{Ri98,Pe99,Ri01}, as
well as  the 2dF Galaxy Redshift survey \cite{Pe02,Ef02},  suggest that
$\Omega = \Omega_{\rm \Lambda} + \Omega_{\rm m} \approx 1$,
or $\Omega_{\rm \Lambda} \approx 0.7$ and $\Omega_{\rm m} \approx 0.3$,
implying the existence of dark energy and dark matter, respectively. One
particular candidate for dark energy is a scalar field usually called 
quintessence \cite{Ca98}. In this way galaxies
are expected to possess dark components and, in accordance with the
rotation curves of stars and gas around the centres of spirals, they
might be in the form of halos \cite{OsPe73}
and  must contribute to at least 3 to 10 times the mass of the visible
matter \cite{KoTu90}.

In order to construct numerical galaxies in a consistent way using gravity
and particle physics components, one must find the correct physical scheme.
The theoretical framework to explain the existence of dark components
finds its origin in theories of elementary particles physics, with the addition 
of the action of gravity.  There exist many theories (grand unification
schemes, string theories, brane worlds, etc) that involve such physics, but
scalar--tensor theories (STT) of gravity are typically found to represent
classical effective descriptions of such original theories \cite{Gr88}.  In
this way, the scalar fields of these theories are the natural candidates to be the
quintessence field \cite{Ca98,Bo00,Am01}, as a remnant of some cosmological
function that contributes  $\Omega_{\rm \Lambda} \approx 0.7$ today.  It has 
been even suggested that the quintessence field is the scalar field that 
also acts on local planetary scales \cite{Fu00} or on galactic 
scales \cite{MaGu01}.  Moreover, massive scalar fields might account to the 
dark matter components of galaxies in the form of halos.

Motivated by the above arguments, we study in the present report some STT
effects in galactic systems.  Specifically, we compute
potential--density pairs coming from such theories in their Newtonian
approximation.  The results presented here will be useful in  
constructing numerical galaxies that are consistent with grand unification 
schemes.

This paper is organised as follows: in section 2 we present the 
Newtonian approximation of a general scalar--tensor theory of gravity.
Solutions are presented in terms of integrals of Green functions, and some
expressions for the velocities and dispersions of stars in galaxies are
given.  In section 3, solutions for a family of potential--density
pairs \cite{De93} and the Navarro, Frenk, and White (NFW) model \cite{Na96-97}
are presented, and some relevant observational quantities are computed. In
section 4 we present conclutions.

\section{Scalar Fields and the Newtonian Approximation}
Let us consider a typical scalar--tensor theory given by the following
Lagrangian
\begin{equation}
{\cal L} = \frac{\sqrt{-g}}{16\pi} \left[ -\phi R + \frac{\omega(\phi)}{\phi}
	(\partial \phi)^2 - V(\phi) \right] + L_M(g_{\mu\nu}) \; ,
\end{equation}
from which we get the gravitational equations,
\begin{eqnarray}
R_{\mu\nu} - \frac{1}{2} g_{\mu\nu} R &=& \frac{1}{\phi}
\left[ 8 \pi T_{\mu\nu} + \frac{1}{2} V g_{\mu\nu}
+ \frac{\omega}{\phi} \partial_\mu \phi \partial_\nu \phi
\right. \nonumber \\
&& \left. -\frac{1}{2} \frac{\omega}{\phi}(\partial \phi)^2 g_{\mu\nu}
+ \phi_{;\mu\nu} - g_{\mu\nu} \, \sq \phi \frac{\mbox{}}{\mbox{}}
\right] \; , \nonumber \\
\end{eqnarray}
and the scalar field equation
\begin{equation}
\sq \phi + \frac{\phi V' - 2V}{3+2\omega} = \frac{1}{3+2\omega} \left[
	8\pi T -\omega' (\partial \phi)^2 \right] \, ,
\end{equation}
where $()' \equiv \frac{\partial }{\partial \phi}$. In the past we have
studied \cite{CeDe95ab}, among others, this type of non--minimal coupling
between gravity and matter fields in a cosmological context in the very early
Universe, where all relativistic effects have to be taken into
account.  In the present study, however, we want to consider the influence
of scalar fields within galaxies, and therefore we need to describe
the theory in its Newtonian approximation, that is, where gravity 
and the scalar fields are weak and velocities of stars are
non--relativistic.  We expect to have small deviations of
the scalar field around the background field, defined here as
$\langle \phi \rangle = G_{N}^{-1} = 1$. If one defines the perturbation
$\bar{\phi} \equiv \phi -1$, the Newtonian approximation
gives \cite{He91}
\begin{eqnarray}
R_{00} = \frac{1}{2} \nabla^2 h_{00} &=& 4\pi \rho
- \frac{1}{2} \nabla^2 \bar{\phi}  \; ,
\label{pares_eq_h00}\\
  \nabla^2 \bar{\phi} - m^2 \bar{\phi} &=& - 8\pi \alpha\rho \; ,
\label{pares_eq_phibar}
\end{eqnarray}
where we have introduced
\begin{displaymath}
\frac{\phi V' - 2V}{3+2\omega} = m^2 \bar{\phi} - m^2 k \bar{\phi}^2
+ \ldots \; ,
\end{displaymath}
and $\alpha \equiv 1 / (3 + 2\omega)$.  In the above expansion we have set
the cosmological constant term equal to zero, since on galactic
scales its influence should be negligible.  We only consider the
influence of dark matter due to the boson field of mass $m$ governed by
Eq.\ (\ref{pares_eq_phibar}), that is a Klein--Gordon equation with a source.

Note that Eq. (\ref{pares_eq_h00}) can be cast as a Poisson equation for
$\psi \equiv (1/2) ( h_{00} + \bar{\phi} )$, 
\begin{equation}
\nabla^2 \psi = 4\pi \rho \; . \label{pares_eq_psi}
\end{equation}
The new Newtonian potential is now given by
\begin{equation}
\Phi_N \equiv \frac{1}{2} h_{00} = \psi - \frac{1}{2} \bar{\phi} \, .
\end{equation}

The next step is to find solutions for this new Newtonian potential given 
a density profile, that is, to find the so--called potential--density pairs. 
First we present the solutions for point--like masses and second  
for general density distributions.

\subsection{Point--like masses}

The solution of these equations is known and given by
\begin{eqnarray} \label{phi00}
\bar{\phi} &=& 2 \alpha u_\lambda \, , \nonumber \\
h_{00} &=& - 2u - 2\alpha u_\lambda \, ,
\end{eqnarray}
where
\begin{eqnarray} \label{uul}
u &=& \sum_s \frac{m_s}{| {\bf r} - {\bf r}_s |} \, , \nonumber \\
u_\lambda &=& \sum_s \frac{m_s}{| {\bf r} - {\bf r}_s |}
{\rm e}^{ -| {\bf r} - {\bf r}_s |/\lambda } \; ,
\end{eqnarray}
with $m_s$ a source mass, and the total gravitational force on a particle of
mass $m_i$ is
\begin{equation}
\sum {\bf F} = -\frac{1}{2} \nabla h_{00} = m_i {\bf a} \, ,
\end{equation}
where $\lambda = \hbar/mc$ is the Compton wavelength of the effective mass ($m$) of
some elementary particle (boson) given through the potential $V(\phi)$ and
$\omega(\phi)$. In what follows we will use $\lambda$ instead of $m^{-1}$.  
This mass can have a range of values depending on particular
particle physics models.  The potential $u$ is the Newtonian part and
$u_\lambda$ is the dark matter contribution which is of the Yukawa type.  
There are two limits: On the one hand, for if 
$r\gg \lambda$ (or $\lambda \rightarrow 0$) 
one recovers the Newtonian theory of gravity. On the other hand, for if
$r \ll \lambda$ (or $\lambda \rightarrow \infty$) one again obtains the  
Newtonian theory, but now with a rescaled Newtonian 
constant, $G \rightarrow G_{N} (1+\alpha)$. There are  
stringent constraints on the possible $\lambda$--$\alpha$ values determined 
by measurements on  local scales \cite{FiTa99}.

In the past the above solutions have been used to solve
the missing mass problem in spirals \cite{Sa84,Ec93} as an alternative to
considering a distribution of dark matter.  This was done assuming that
most of the galactic mass is located in the galactic centre, and then  
considering the centre to be a point source.  In our present investigation we 
do not avoid dark matter, since our model predicts that bosonic dark matter
produces, through a scalar field associated to it, a modification of 
Newtonian gravity theory.  This dark matter is presumably clumped in the
form of dark halos.   Therefore we will consider in what follows that
a dark halo is spherically distributed along an observable spiral and
beyond, having some density profile (section 3).  Next, we compute the
potentials, and some astrophysical quantities, for general halo density
distributions.

\subsection{General density distributions}

General solutions to Eqs. (\ref{pares_eq_phibar}) and (\ref{pares_eq_psi})
can be found in terms of the corresponding Green functions
\begin{eqnarray}
\psi &=& -\int d{\bf r}_s \frac{\rho({\bf r}_s)}{|{\bf r}-{\bf r}_s|}
+ \mbox{B.C.} \label{pares_eq_gralpsi} \; , \\
\bar{\phi} &=& 2\alpha\int d{\bf r}_s
\frac{\rho({\bf r}_s)  {\rm e}^{- |{\bf r}-{\bf r}_s|/\lambda}}
	{| {\bf r}-{\bf r}_s|} + \mbox{B.C.} \label{pares_eq_gralphi} \; ,
\end{eqnarray}
and the new Newtonian potential is
\begin{eqnarray}
\Phi_N &=&  \psi
- \frac{1}{2} \bar{\phi} \nonumber \\
&=& -\int d{\bf r}_s
\frac{\rho({\bf r}_s)}{|{\bf r}-{\bf r}_s|} \nonumber \\
&& -\alpha\int d{\bf r}_s \frac{\rho({\bf r}_s)
{\rm e}^{- |{\bf r}-{\bf r}_s|/\lambda}}
{| {\bf r}-{\bf r}_s|} + \mbox{B.C.} \label{pares_eq_gralPsiN}
\end{eqnarray}
The first term of Eq. (\ref{pares_eq_gralPsiN}), given by $\psi$, is the
contribution of the usual Newtonian gravitation (without scalar
fields), while information about the scalar field is contained in the
second term, that is, arising from the influence function determined by the
Klein--Gordon Green function, where the coupling $\omega$ ($\alpha$) enters
as part of a source factor.

Given that we are interested in spherical galaxies and bulges in what follows
we will only consider the case of spherical symmetry. Additionally, we use
flatness boundary conditions (B.C.) at infinity, such that the boundary terms 
in Eqs. (\ref{pares_eq_gralpsi})-(\ref{pares_eq_gralphi}) are zero.  Moreover,
regularity conditions must be applied to spatial points where the potentials
are singular. For spherical systems these conditions mean that
$d \psi/dr=d \bar{\phi}/dr=0$ at the origin.  Accordingly, performing the
integrals in these equations, we obtain $\psi$ for a system of radius $R$, 
\begin{equation}\label{pares_eq_finalpsi}
\psi(r_a) = -\frac{M}{a}\left\{ \begin{array}{ll}
\frac{m(r_a)}{r_a} + o(r_a) & ; \; r_a < R_a \\[0.05in]
\frac{m(R_a)}{r_a} & ; \; r_a \ge R_a
\end{array} \right. \; ,
\end{equation}
and for $\bar{\phi}$
\begin{equation}\label{pares_eq_finalphi}
\bar{\phi}(r_a) = \frac{2M}{a} \eta \times \left\{ \begin{array}{ll}
\frac{{\rm e}^{-r_a/\lambda_a}}{r_a} p(r_a)
+ 
\frac{\sinh(r_a/\lambda_a)}{r_a} q(r_a)  & \mbox{\hspace{-0.1in}}  ; 
r_a < R_a \\ \\
\frac{{\rm e}^{-r_a/\lambda_a}}{r_a} p(R_a)  & \mbox{\hspace{-0.1in}} 
;  r_a \ge R_a \, ,
\end{array} \right.
\end{equation}
where we have defined the functions
\begin{eqnarray}
m(r_a) &\equiv& \int_0^{r_a} dx \; x^2 \tilde{\rho}(x) \; , \nonumber \\
o(r_a) &\equiv&  \int_{r_a}^{R_a} dx \; x \tilde{\rho}(x) \; , \nonumber \\
p(r_a) &\equiv&  \int_0^{r_a} dx \; x^{2} \frac{\sinh(x/\lambda_a)}{x/\lambda_a} 
\tilde{\rho}(x) \; , \\
q(r_a) &\equiv&  \int_{r_a}^{R_a} dx \; x^{2} \frac{\exp(-x/\lambda_a)}{x/\lambda_a} 
\tilde{\rho}(x) \; , \nonumber
\end{eqnarray}
and we have written the density as
\begin{equation}
\rho(r) =  \frac{M}{4\pi a^3} \tilde{\rho}(r_a)  \; .
\end{equation}
Here $M$ is a reference or scaling mass and $a$ is a scaling length that 
we use to define the following quantities: $r_a \equiv r/a$, $R_a \equiv R/a$, 
$\lambda_a \equiv \lambda/a$, and  $\eta \equiv \alpha/ \lambda_a$.  Then, the 
new Newtonian potential can be  written as
\begin{equation}\label{pares_eq_finalPhiN}
\Phi_N(r_a) = 
-\frac{M}{a}  F(r_a) \, ,
\end{equation}
where
\begin{equation}\label{pares_eq_F}
F(r_a) =  \left\{ \begin{array}{ll}
\frac{m(r_a)}{r_a} + o(r_a) + \eta \, 
\left[ \frac{{\rm e}^{-r_a/\lambda_a}}{r_a/\lambda_a}  p(r_a) \right. & \\
\left. + \frac{\sinh(r_a/\lambda_a)}{r_a/\lambda_a} q(r_a) \right]  
& \mbox{\hspace{-0.1in}} ;  r_a < R_a  \\ \\
\frac{m(R_a)}{r_a} + \eta \, 
\frac{{\rm e}^{-r_a/\lambda_a}}{r_a/\lambda_a} p(R_a)  & 
\mbox{\hspace{-0.1in}} ;  r_a \ge R_a \, .
\end{array} \right.
\end{equation}

\subsection{Intrinsic and projected properties}

We now want to present the intrinsic and observable quantities of
interest that characterize the steady state of spherical galaxies and
bulges. These include the circular velocity, velocity dispersion, distribution 
function, projected velocity dispersion, and projected density.

The circular velocity for a spherical galactic system is, from the above 
equations, 
\begin{eqnarray} \label{pares_eq_vc2_1}
v_c^2 &=& r \frac{d\Phi_N}{dr} 
= - \frac{M}{a} \, r_{a} \, \frac{dF(r_a)}{dr_a}  \, .
\end{eqnarray}
%
For  $r_a < R_a$ one obtains straightforwardly the circular velocities 
for stars inside a dark halo of size $R$.   For $r_a\ge R_a$ one gets 
\begin{eqnarray} \label{v2-plm}
\frac{a}{M} v_c^2 &=&
\frac{m(R_a)}{r_a} + \alpha
\left[ o(R_a) \sinh(R_a/\lambda_a) - Q(R_a) \right] \nonumber \\
&& \times \frac{{\rm e}^{-r_a/\lambda_a}}{r_a/\lambda_a} (1 +\frac{r_a}{\lambda_a}) \, ,
\end{eqnarray}
where 
\begin{equation}
Q(r_a) \equiv  \frac{1}{\lambda_a} \, \int_0^{r_a} dx \; x 
\; o(x) \; \cosh(x/\lambda_a)  \, ,
\end{equation}
which gives the circular velocities outside the halo (if there are 
stars there).  From Eqs. (\ref{pares_eq_F}) and (\ref{pares_eq_vc2_1}) one 
obtains, after  redefinition of the multiplicative constant, the expression 
found in  the past for rotational velocities due to point masses \cite{Sa84}.  

\bigskip

A full description of stars in a galaxy is given by the collisionless 
Boltzmann equation, which after being integrated over all  velocities yields    
the Jeans equations relating the potential and density of a system 
to its velocity dispersion \cite{BiTr94}. In particular, the velocity 
dispersion profile is important in determining the stationary state of 
the system.  For equilibrium stationary states we found the velocity 
dispersion for a spherically symmetric stellar 
system, using  Eq.\ (\ref{pares_eq_vc2_1}),  to be 
\begin{eqnarray}
\bar{v_r^2} = - \frac{M}{a} 
\frac{1}{\tilde{\rho}(r_a)}
\int_{r_a}^\infty dx\;  \tilde{\rho}(x) 
\frac{dF(x)}{dx}  \, ,
\label{pares_eq_vr}
\end{eqnarray}
where we have assumed that $\bar{v_r^2}=\bar{v_\theta^2}$ and
$\bar{v_r}=\bar{v_\theta}=0$.   We have used similar expressions to generate
initial conditions for proto--galaxies in order to study the influence of scalar fields
on the transfer of angular momentum during collisions \cite{Ro01}.

The distribution function of the isotropic models can be calculated using
the Abel transform \cite{BiTr94}. First, we define ${\cal E}=-(a/M)E$,
$E$ the total energy, and $\Psi(r_a)=-(a/M) \Phi_N(r_a)$, and  
we obtain $r$ as a function of $\Psi$ by inverting. The mass density is
\begin{equation}
\rho(r)=4\pi\left(\frac{M}{a}\right)^{3/2}
\int_0^\Psi f({\cal E}) \sqrt{2(\Psi-{\cal E})} d{\cal E}  \, .
\end{equation}
The inversion of this equation is done using the Eddington formula and
we obtain the distribution function
\begin{equation}
f({\cal E})=\frac{1}{\sqrt{8}\pi^2}\left(\frac{a}{M}\right)^{3/2}
\left[
\int_0^{\cal E} \frac{d^2 \rho}{d\Psi^2} \frac{d\Psi}{\sqrt{{\cal E}-\Psi}}
+\frac{1}{\sqrt{{\cal E}}} \left(\frac{d\rho}{d\Psi}\right)_{\Psi=0}
\right] .
\end{equation}

The projected velocity dispersion for the isotropic system is given by 
\cite{BiTr94}
\begin{equation}
\sigma_p^2(R)=\frac{2}{\Sigma(R)}
\int_R^\infty \rho(r) \bar{v_r^2} \frac{r\; dr}{\sqrt{r^2-R^2}} \, ,
\end{equation}
where $\Sigma(R)=2\int_R^\infty dr \; \rho(r)r/\sqrt{r^2-R^2}$
is the projected density. This also gives us the cumulative surface density
\begin{equation}
S(R)=2\pi\int_0^R \Sigma(R) R dR \, .
\end{equation}


\section{Potential--density pairs}

In the past some potential--density pairs have been studied within the
framework of Newtonian gravity.  The Jaffe  \cite{Ja83} and
Hernquist density models \cite{He90} have been generalised by 
Dehnen \shortcite{De93} by introducing a free parameter ($\gamma$) 
that determines particular density mo\-dels for galaxy 
description. Another density profile of interest, obtained from
N-body cosmological simulations, has been proposed by 
NFW \cite{Na96-97}. In this section we will consider both models 
in order to study the influence of scalar fields on spherical systems.

\subsection{Dehnen's profile}

We use the family of density profiles for spherical halos and
bulges of galaxies proposed by Dehnen \shortcite{De93}, 
\begin{equation}\label{pares_eq_wdrho}
\rho(r) = \frac{(3-\gamma)M}{4\pi} \frac{a}{r^\gamma(r+a)^{4-\gamma}} \, ,
\end{equation}
where $a$ is a scaling radius and $M$ denotes the total mass. The Hernquist
profile corresponds to $\gamma=1$ and Jaffe's to $\gamma=2$.

Solving the Poisson equation without a scalar field, {\it i.e.} for
Newtonian gravity, the potential that corresponds to the density of
Eq. (\ref{pares_eq_wdrho}) is \cite{De93}
\begin{eqnarray}\label{pares_eq_wdpot}
\Phi_N=\psi = \frac{M}{a} \times
\left\{ \begin{array}{ll}
-\frac{1}{2-\gamma} \left[ 1
- \left(\frac{r}{r+a}\right)^{2-\gamma} \right]
	& ; \; \gamma \ne 2 \\
		\ln\frac{r}{r+a} & ; \; \gamma=2 \; .
	\end{array} \right.
\end{eqnarray}
Due to the influence of the scalar field, however, $\psi(r)$ does not
represent the total gravitational potential of the system.  The effective
Newtonian potential is now $\Phi_N=\frac{1}{2}h_{00}(r)$ determined by
Eq. (\ref{pares_eq_gralPsiN}). Thus, using Eq. (\ref{pares_eq_wdrho}) for
the density, we obtain for $r_a<R_a$
\begin{eqnarray}
\frac{a}{M}\Phi_N &=&
-\int_{r_a}^{R_a} dx \frac{m(x)}{x^2}  \label{pares_eq_DehnenPhi}\\
&& - \eta \left[ \frac{{\rm e}^{-r_a/\lambda_a}}{r_a/\lambda_a} p(r_a) +
\frac{\sinh(r_a/\lambda_a)}{r_a/\lambda_a} q(r_a) \right] \nonumber \; ,
\end{eqnarray}
and for $r_a\ge R_a$
\begin{eqnarray}
\frac{a}{M}\Phi_N = -\frac{m(R_a)}{r_a}
  - \eta \frac{{\rm e}^{-r_a/\lambda_a}}{r_a/\lambda_a} p(R_a)  \, .
\end{eqnarray}

Using Eq. (\ref{pares_eq_wdrho}) one finds explicitly 
the functions  $m$, $p$, and $q$:
\begin{eqnarray}
m(r_a) \mbox{\hspace{-0.1in}} &=&  \mbox{\hspace{-0.1in}} 
(3-\gamma)\int_0^{r_a} dx \frac{x^{2-\gamma}}{(x+1)^{4-\gamma}} 
=\left(\frac{r_a}{r_a+1}\right)^{3-\gamma}\; , \\
p(r_a) \mbox{\hspace{-0.1in}} &=& \mbox{\hspace{-0.1in}} 
(3-\gamma) \lambda_a \int_0^{r_a} dx 
\frac{x^{1-\gamma}}{(x+1)^{4-\gamma}} 
\sinh(x/\lambda_a)  \\
&=& \mbox{\hspace{-0.1in}}  (3-\gamma) \left[ 
\frac{1}{3-\gamma} r_a^{3-\gamma} 
\mbox{}_2F_1(3-\gamma,4-\gamma;4-\gamma,-r_a)
\right. \nonumber 
\\
&& \left. + \frac{1}{6\lambda_a^2} \frac{1}{5-\gamma} 
r_a^{5-\gamma} 
\mbox{}_2F_1(5-\gamma,4-\gamma;6-\gamma,-r_a)
+ 
\ldots
\right] \; , \nonumber \\
q(r_a) \mbox{\hspace{-0.1in}} &=& \mbox{\hspace{-0.1in}} 
(3-\gamma) \lambda_a \int_{r_a}^{R_a} dx
\frac{x^{1-\gamma}}{(x+1)^{4-\gamma}} {\rm e}^{-x/\lambda_a}
\\
&=& \mbox{\hspace{-0.1in}} (3-\gamma) 
\left\{
\left[
\frac{\lambda_a}{2-\gamma} R_a^{2-\gamma} 
\mbox{}_2F_1(2-\gamma,4-\gamma;3-\gamma,-R_a)
\right.\right. 
\nonumber \\
&& \mbox{\hspace{-0.1in}}  -  
\frac{1}{3-\gamma} R_a^{3-\gamma} 
\mbox{}_2F_1(3-\gamma,4-\gamma;4-\gamma,-R_a)
\nonumber \\
&& 
\mbox{\hspace{-0.1in}}  \left. + \frac{1}{2\lambda_a} 
\frac{1}{4-\gamma} R_a^{4-\gamma} 
\mbox{}_2F_1(4-\gamma,4-\gamma;5-\gamma,-R_a)
- \ldots
\right] 
\nonumber \\
&&\mbox{\hspace{-0.1in}}  -\left[
\frac{\lambda_a}{2-\gamma} 
r_a^{2-\gamma} 
\mbox{}_2F_1(2-\gamma,4-\gamma;3-\gamma,-r_a)
\right. 
\nonumber \\
&& \mbox{\hspace{-0.1in}}  -  
\frac{1}{3-\gamma} r_a^{3-\gamma} 
\mbox{}_2F_1(3-\gamma,4-\gamma;4-\gamma,-r_a)
\nonumber \\
&& 
\mbox{\hspace{-0.1in}}  
\left. \left. + \frac{1}{2\lambda_a} 
\frac{1}{4-\gamma} r_a^{4-\gamma} 
\mbox{}_2F_1(4-\gamma,4-\gamma;5-\gamma,-r_a)
- \ldots
\right] 
\right\} \; , \nonumber
\end{eqnarray}
where $\mbox{}_2F_1(a,b;c,z)$ is a hypergeometric function.

For $r_a < R_a$ the integral in the first term of 
Eq.\ (\ref{pares_eq_DehnenPhi}) can be evaluated,
giving for the standard  Newtonian potential
\begin{eqnarray}
\psi=-\frac{M}{a}  \times
\left\{ \begin{array}{ll}
\frac{1}{2-\gamma} \left[ 
\left(\frac{R_a}{R_{a} +1}\right)^{2-\gamma}
- \left(\frac{r_a}{r_{a} +1}\right)^{2-\gamma} \right]
	& ; \quad \gamma \ne 2 \\[0.1in]
		\ln\frac{R_a}{R_{a} +1}
	-\ln\frac{r_a}{r_{a} +1} & ; 
\quad \gamma=2,
	\end{array} \right.
\end{eqnarray}
which for $R_a \rightarrow \infty$ reduces to 
Eq.\ (\ref{pares_eq_wdpot}).

The circular velocity is, for $r_a< R_a$,
\begin{eqnarray}\label{pares_eq_circular_a}
\frac{a}{M}v_c^2 &=&
\frac{m(r_a)}{r_a}
  +  \eta \,  \left[
\left(1+\frac{r_a}{\lambda_a}\right)
\frac{{\rm e}^{-r_a/\lambda_a}}{r_a/\lambda_a}
\; p(r_a) \right.  \\
&&+ \left.
\left(1
-\frac{r_a}{\lambda_a}\coth(r_a/\lambda_a)\right) \;
\frac{\sinh(r_a/\lambda_a)}{r_a/\lambda_a} q(r_a) \right] 
\nonumber 
\; ,  
\end{eqnarray}
and for $r_a\ge R_a$
\begin{eqnarray}\label{pares_eq_circular_b}
\frac{a}{M}v_c^2 =
\frac{m(R_a)}{r_a} 
+ \eta \,
\left(1+\frac{r_a}{\lambda_a}\right)
\frac{{\rm e}^{-r_a/\lambda_a}}{r_a/\lambda_a}
\; p(R_a) \; .
\end{eqnarray}
The first right--hand--side term in both Eqs. (\ref{pares_eq_circular_a}) 
and (\ref{pares_eq_circular_b}) is the contribution 
from $\psi$ and their second term is the contribution from the 
scalar field. The latter term depends on  $R_a$, $\lambda_a$ and 
$\eta= \alpha/\lambda_a$, rather than on $\alpha$ alone,
which is typical for forces exerted on point  
systems \cite{FiTa99}.  Therefore, the stringent constraints on  
$\alpha$ found for point  masses at local scales are less severe
here, since the above terms can still yield a large amplitude, even when  
$\alpha$  is small.  One sees that for $r \gg R,\,  \lambda$, the influence 
of scalar field decays exponentially, recovering the standard
Newtonian result.

The velocity dispersion can be computed by substituting
Eqs. (\ref{pares_eq_circular_a}) and (\ref{pares_eq_circular_b}) 
into Eq. (\ref{pares_eq_vr}), and this determines
completely the initial conditions for spherical galaxies.  Fig. 1 shows
the circular velocity and the velocity dispersion curves 
($s_r=\frac{a}{M}\bar{v_r^2}$) for the Hernquist model ($\gamma=1$) for  
$\eta=1$ (upper curves) and $\eta=0$ (lower curves), 
i.e., the upper curves are computed with the scalar field taken into account  
and the lower curves without a scalar field.  In the limit
$\lambda \rightarrow \infty$ one recovers the Newtonian result with a modified $G$ as 
expected.  The computations were done with $R \rightarrow \infty$.

\begin{figure}
\epsfxsize=20pc  \epsfbox{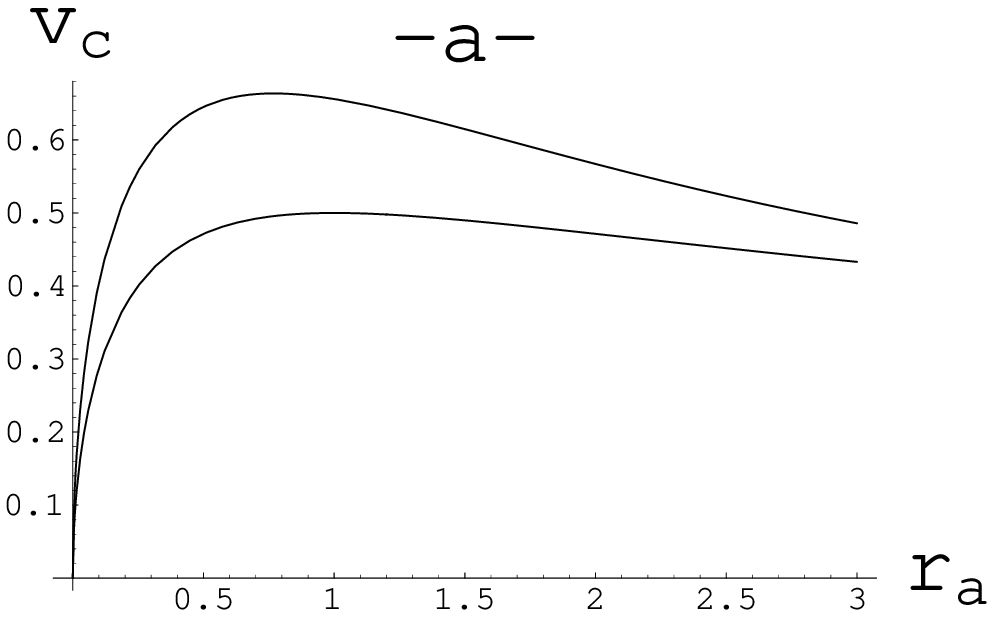}
\epsfxsize=20pc  \epsfbox{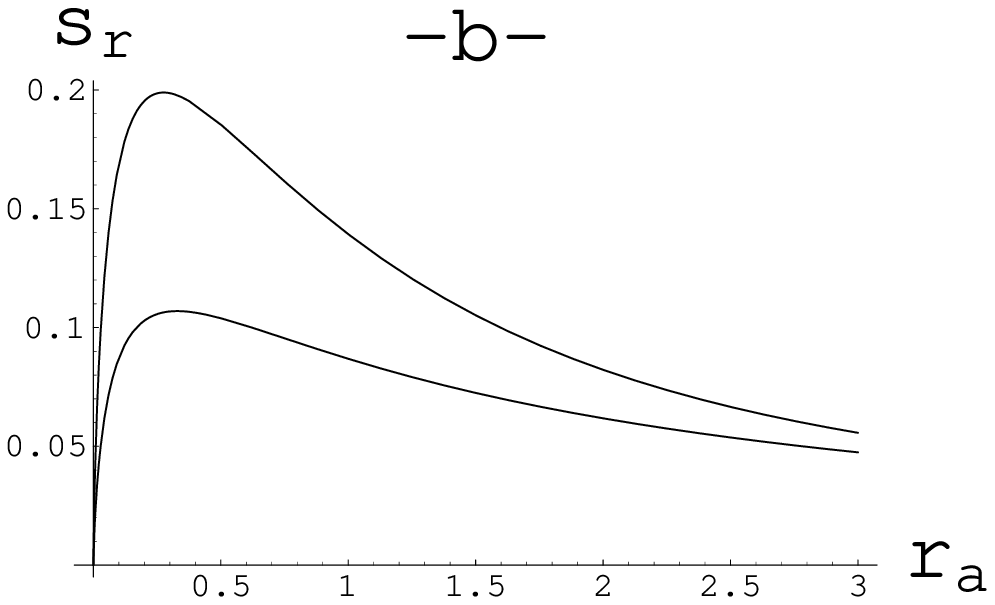}
\caption{Rotation curves (a) and velocity dispersion curves (b) 
for the Hernquist model. The upper curves were computed 
with $\lambda_a=1$ and $\eta=1$, while the 
lower curves were obtained for $\lambda_a=1$ and
$\eta=0$, i.e., without a scalar field.}\label{pares_fig_01}
\end{figure}

\subsection{NFW profile}

$N$-body cosmological simulations \cite{Na96-97} suggest a universal density
profile for spherical dark halos of the following type:
\begin{equation}\label{pares_eq_nfwrho}
\rho(r) = \frac{\rho_{\rm crit} \, \delta_c}{(r/r_{s}) \, (1+r/r_{s})^{2}} \, ,
\end{equation}
where $r_{s}$ is a scale radius, $\delta_c$ is a dimensionless constant, and
$\rho_{\rm crit}$ is the critical density (for closure) of the Universe. This
profile is the same as Dehnen's with $\gamma=1$ (Hernquist model) for 
$r \ll r_s$, but it differs at other radii.

This halo density profile was obtained for cosmological simulations
in which the Newtonian theory of gravity was used.  Because we are
considering $\lambda \ll H^{-1}$, then all predictions of such
cosmological simulations must be taken into account; this is because
in the present formalism for $r\gg \lambda$ one gets the Newtonian theory.

Solving Eqs.\ (\ref{pares_eq_phibar}) and (\ref{pares_eq_psi}) with 
the density profile (\ref{pares_eq_nfwrho}), we get for $r_a < R_a$,  
\begin{eqnarray}
\frac{a}{M} \Phi_N(r_a) \mbox{\hspace{-0.05in}} 
&=& 
-\frac{m(r_a)}{r_a}-o(r_a)  \nonumber \\
&&-\eta \left[ 
\frac{e^{-r_a/\lambda_a}}{r_a/\lambda_a} 
p(r_a)
+\frac{\sinh(r_a/\lambda_a)}{r_a/\lambda_a} q(r_a) \right]
\end{eqnarray}
and for $r_a\ge R_a$
\begin{eqnarray}
\frac{a}{M} \Phi_N(r_a) = -\frac{m(R_a)}{r_a} 
-\eta 
\frac{e^{-r_a/\lambda_a}}{r_a/\lambda_a} p(R_a) , 
\end{eqnarray}
where the reference length $a=r_s$. 
The circular velocity is given 
by Eqs.\ (\ref{pares_eq_circular_a}) 
and 
(\ref{pares_eq_circular_b})
where the functions $m(r_a)$, $o(r_a)$, 
$p(r_a)$, and $q(r_a)$
are given 
by
\begin{equation}
m(r_a)=\frac{1}{1+r_a}+\ln(1+r_a)-1 \; 
,
\end{equation}
\begin{equation}
o(r_a)=\frac{1}{1+r_a}-\frac{1}{1+R_a} 
\; ,
\end{equation}
\begin{eqnarray}
p(r_a) \mbox{\hspace{-0.1in}} 
&=&
-\frac{\lambda_a}{1+r_a}\sinh(r_a/\lambda_a) \nonumber \\
&& 
+ 
\cosh(1/\lambda_a)
\left[
\mbox{Chi}\left(\frac{1+r_a}{\lambda_a}\right)
-\mbox{Chi}\left(\frac{1}{\lambda_a}\right)
\right] 
\nonumber \\
&& 
-
\sinh(1/\lambda_a)
\left[
\mbox{Shi}\left(\frac{1+r_a}{\lambda_a}\right)
-\mbox{Shi}\left(\frac{1}{\lambda_a}\right)
\right] ,
\end{eqnarray}

\begin{eqnarray}
q(r_a) \mbox{\hspace{-0.1in}} 
&=&
  \lambda_a \, \frac{{\rm e}^{-r_a/\lambda_a}}{1+r_a}
- \lambda_a \, \frac{{\rm e}^{-R_a/\lambda_a}}{1+R_a}
\nonumber \\ 
&&+{\rm e}^{1/\lambda_a}
\left[
\mbox{Ei}\left(-\frac{1+r_a}{\lambda_a}\right)
-\mbox{Ei}\left(-\frac{1+R_a}{\lambda_a}\right)
\right] \, ,
\end{eqnarray}
where $\mbox{Ei}(z)$, $\mbox{Shi}(z)$, and 
$\mbox{Chi}(z)$
are the exponential, hyperbolic sine and hyperbolic 
cosine integrals, respectively.

\begin{figure}
\epsfxsize=20pc  \epsfbox{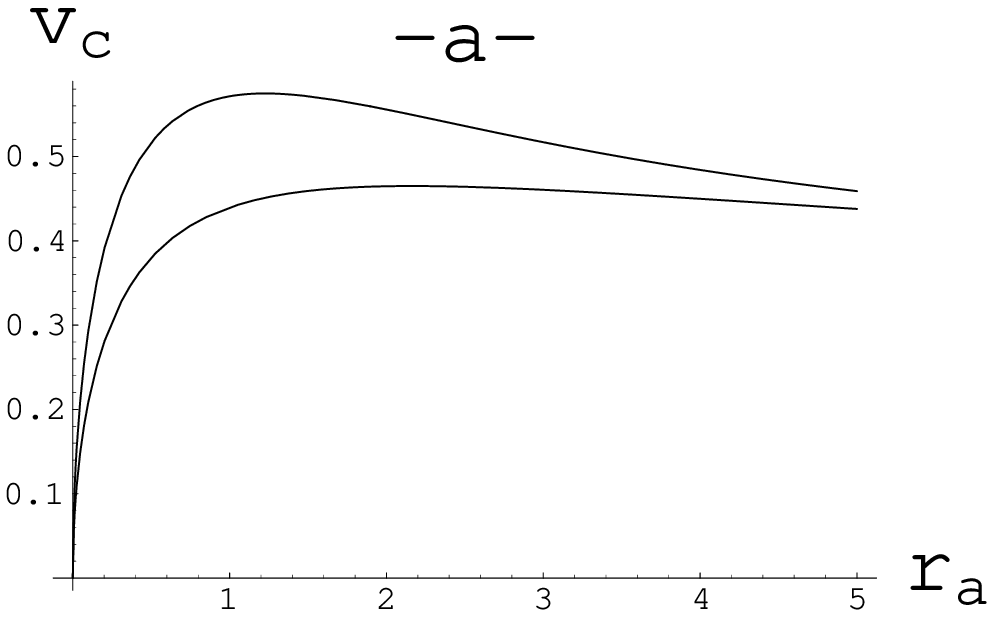}
\epsfxsize=20pc  \epsfbox{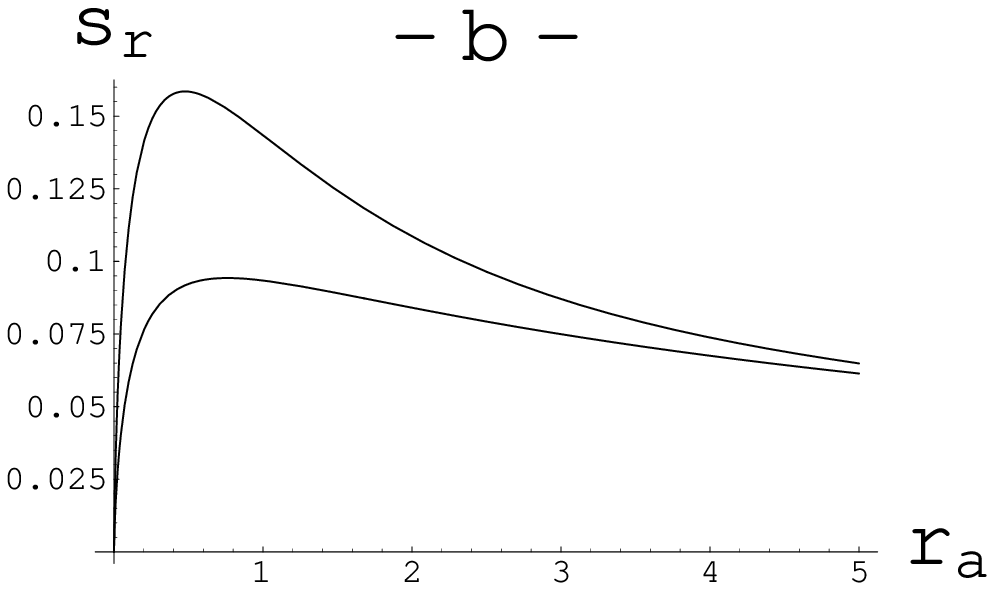}
\caption{Rotation curves (a) and velocity dispersion curves (b) for 
the NFW density profile. $\lambda_a =1$ and upper (lower) curve was 
computed with $\eta=1$ ($\eta=0$).}\label{pares_fig_02}
\end{figure}

In Fig.\ 2 we plot the circular velocity and its dispersion for the NFW profile. The 
values of $\lambda_a$ and $\eta$ are the same as in Fig.\ 1.  The upper curves 
correspond to the case in which we take into account the scalar 
field, while the lower curves were computed for standard  
Newtonian gravity.  We took the size of the distribution of mass to be  $R=20\; r_s$.   
\bigskip 

The influence of the scalar field is to enhance, especially in the inner regions, both 
the velocity curves and dispersion.  These effects are a little more pronounced in the 
Hernquist model than in the NFW model.

\section{Discussion and Conclusions}
We have found potential--density pairs of spherical galactic systems within
the context of linearised scalar--tensor theories of gravitation.  The
influence of massive scalar fields is given by $\bar{\phi}$, determined by
Eq. (\ref{pares_eq_finalphi}). General expressions have been given for
circular velocities and dispersions of {\it stars} in the spherical system; by stars
we mean probe particles that follow the dark halo potential.  
Specifically, these results were used to find potential--density pairs
for the generic Dehnen density parametrisation, as well as for the NFW
profile.  In general the contribution due to massive scalar fields is 
non--trivial, see for instance 
Eqs. (\ref{pares_eq_circular_a}-\ref{pares_eq_circular_b}), and interestingly, forces 
on circular orbits of stars depend on the amplitude terms $R_a$, $\lambda_a$ and 
$\eta= \alpha / \lambda_a$.  This means that even when local 
experiments force $\alpha$ to be a very small number \cite{FiTa99}, the 
amplitude of forces exerted on stars is not necessarily very small and may 
contribute significantly to the dynamics of stars.  Alternatively, one may interpret 
the local Newtonian constant as given by $ (1 + \alpha) \langle \phi \rangle^{-1} $, 
instead of being given by $\langle \phi \rangle^{-1}$ ($1$ in our convention). In this case,
the local measurement constraints are automatically satisfied, and at  
scales larger than $\lambda$ one sees a reduction of $1/(1+\alpha)$ in the 
Newtonian constant. If this were the case, then the upper curves in figures 1 and 2 
would have to be multiplied by $1/(1+\alpha)$.
\bigskip 

In the past, different authors have used  point solutions, Eqs. (\ref{phi00}) 
and (\ref{uul}), to solve the missing 
mass problem encountered in the rotational velocities of spirals
and in galaxy cluster dynamics \cite{Sa84,Ec93}. These models were used 
as an alternative to avoid dark matter. 
Indeed, for single galaxies one can adjust the
parameters ($\alpha$, $\lambda$) to solve these problems without
the need for dark matter.  However, these models do not provide a good 
description of the systematics of galaxy rotational curves because they 
predict the scale $\lambda$ to be independent of the galactic luminosity, and this conflicts
with observations for different galactic sizes \cite{Ag01} unless one assumes
various $\lambda$'s, and hence various fundamental masses, $m$, one for each
galaxy size. Such a particle spectrum is not expected from theoretical 
arguments; it represents a considerable fine tuning of masses.  This criticism  
would also apply to our models.  The purpose of the 
present investigation was, however, not to present a model alternative 
to dark matter in order to solve the missing mass problem, but to compute the influence 
of scalar--field dark  matter distributed in the form of a dark halo.  This
contribution, together with bulges and disks, will give rise to a flat 
velocity curve, as in Newtonian mechanics. Yet, the influence
of the scalar field from STT yields some profile modification, especially in 
inner regions, as is shown in  figs. 1 and 2.

\vspace {1 cm}
{\Large\bf Acknowledgments}
\bigskip

This work was supported in part by the DFG, DAAD and CONACYT grant number 33278-E.


\end{document}